\documentclass[aps,pra,twocolumn,amsmath,amssymb,nofootinbib,showpacs,superscriptaddress]{revtex4-1}
\usepackage[english]{babel}
\usepackage{latexsym}
\usepackage{graphics}
\usepackage{graphicx}
\usepackage{epsfig}
\usepackage{color}
\usepackage{bm}
\usepackage{amsmath}
\usepackage{amssymb}
\usepackage{amsthm}
\usepackage{dcolumn}
\usepackage{bm}
\usepackage{float}
\usepackage{hyperref}
\usepackage{color}
\usepackage{epstopdf}
\usepackage{cleveref}
\usepackage[svgnames]{xcolor}
\hypersetup{hidelinks,colorlinks=true,allcolors=DarkBlue}
\usepackage{enumerate}

\begin{document}

\preprint{APS/123-QED}

\title{Quantum-secured data transmission in urban fibre-optic communication lines}

\author{A.V.~Duplinskiy}
\affiliation{Russian Quantum Center, Skolkovo, Moscow 143025, Russia}
\affiliation{Moscow Institute of Physics and Technology, Dolgoprudny, Moscow Region 141700, Russia}
\author{E.O.~Kiktenko}
\affiliation{Russian Quantum Center, Skolkovo, Moscow 143025, Russia}
\affiliation{QApp, Skolkovo, Moscow 143025, Russia}
\affiliation{Steklov Mathematical Institute of Russian Academy of Sciences, Moscow 119991, Russia}
\author{N.O.~Pozhar}
\affiliation{Russian Quantum Center, Skolkovo, Moscow 143025, Russia}
\affiliation{QApp, Skolkovo, Moscow 143025, Russia}
\author{M.N.~Anufriev}
\affiliation{Russian Quantum Center, Skolkovo, Moscow 143025, Russia}
\affiliation{QApp, Skolkovo, Moscow 143025, Russia}
\author{R.P.~Ermakov}
\affiliation{Russian Quantum Center, Skolkovo, Moscow 143025, Russia}
\author{A.I.~Kotov}
\affiliation{AMICON Co., Ltd., Moscow 117587, Russia}
\author{A.V.~Brodskiy}
\affiliation{Sberbank of Russia, Moscow 117997, Russia}
\author{R.R.~Yunusov}
\affiliation{Russian Quantum Center, Skolkovo, Moscow 143025, Russia}
\author{V.L.~Kurochkin}
\affiliation{Russian Quantum Center, Skolkovo, Moscow 143025, Russia}
\affiliation{QRate, Skolkovo, Moscow 143025, Russia}
\author{A.K.~Fedorov}\email{akf@rqc.ru}
\affiliation{Russian Quantum Center, Skolkovo, Moscow 143025, Russia}
\affiliation{QApp, Skolkovo, Moscow 143025, Russia}
\affiliation{QRate, Skolkovo, Moscow 143025, Russia}
\author{Y.V.~Kurochkin}
\affiliation{Russian Quantum Center, Skolkovo, Moscow 143025, Russia}
\affiliation{QRate, Skolkovo, Moscow 143025, Russia}

\begin{abstract}
Quantum key distribution (QKD) provides information-theoretic security in communications based on the laws of quantum physics.
In this work, we report an implementation of quantum-secured data transmission in the infrastructure of Sberbank of Russia in standard communication lines in Moscow
The experiment is realized on the basis of the already deployed urban fibre-optic communication channels with significant losses.
We realize the decoy-state BB84 QKD protocol using the one-way scheme with polarization encoding for generating keys.
Quantum-generated keys are then used for continuous key renewal in the hardware devices for establishing a quantum-secured VPN Tunnel between two offices of Sberbank. 
The used hybrid approach offers possibilities for long-term protection of the transmitted data, and it is promising for integrating into the already existing information security infrastructure. 
\end{abstract}

\maketitle

\section{Introduction} 

Recent progress in creating quantum algorithms posses a serious threat on the central element of currently used tools for ensuring information security, the key distribution infrastructure. 
The majority of methods for key distribution is based on the assumption of the computational complexity of several mathematical tasks, 
such as large number factorization~\cite{Schneier}.
However, Shor's algorithm for a quantum computer allows solving these problems in a polynomial time~\cite{Shor1997}. 
Moreover, absence of an efficient classical (non-quantum) algorithm breaking such public-key cryptosystems still remains unproved.

Quantum computers have less of an effect on symmetric cryptographic primitives, such as GOST block cipher if it is assumed that the master key has been distributed secretly, 
since Shor's algorithm does not apply, and then exponential speedups are not expected~\cite{Shor1997}. 
Nevertheless, Grover's search algorithm~\cite{Grover1996} would allow quantum computers a quadratic speedup in brute force search, 
which means that the key management in terms of the key size and the key refresh time for such primitives needs to be reconsidered. 

An ultimate and practical solution for the key distribution problem is the QKD technology. 
The QKD method uses the possibility to encode information in states of single photons, transmit them through optical channels, and measure on the receiver side~\cite{Gisin2002,Lo2015,Lo2016}.
By virtue of a number quantum-mechanical phenomena, this allows one to exclude possibilities for undetectable eavesdropping~\cite{Gisin2002}.
It is important to note that the method for preparation and measurements of quantum states, so-called QKD protocol, 
should guarantee the absence of undetectable eavesdropping.
Presently, decoy-state BB84 QKD is a standard technique, 
which provides security and significant key rates for a large distance between parties~\cite{Hwang2003,Lo2005,Wang2005,Ma2005,Curty2014,Lim2014,Ma2017,Trushechkin2017}.

\begin{figure}[t]
\begin{centering}
\includegraphics[width=1\columnwidth]{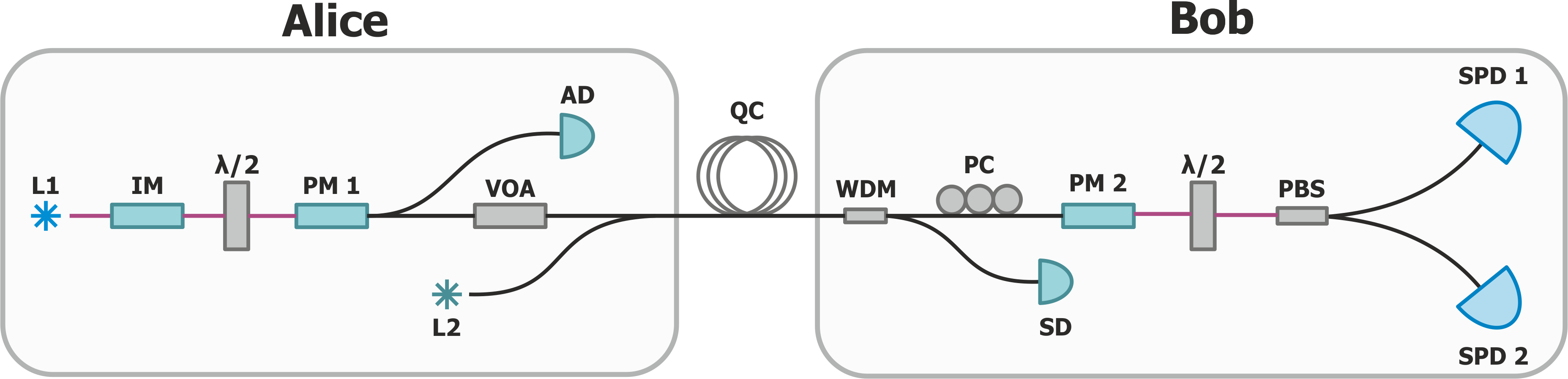}
\end{centering}
\vskip -7mm
\caption
{Setup for QKD using the polarization-encoding scheme with the light source L1, 
intensity modulator IM, 
half-wave plate $\lambda/2$, 
phase modulators PM1 and PM2, variable optical attenuator VOA, 
synchronization laser L2, 
analyzing detector AD, 
wavelength-division multiplexing filter WDM, 
polarization controller PC, 
synchronization detector SD, 
quantum channel (urban fiber-optics channel) QC, 
polarization beam splitter PBS, and single-photon detectors SPD1 and SPD2. 
The polarization maintaining fiber is used for connections between L1 and PM1 for Alice, and between PM2 and PBS for Bob.}
\label{fig:setup}
\end{figure}

In this work, we report the experimental demonstration of quantum-secured data transmission in standard communication lines in Moscow.
Due to significant losses in the urban fibre-optic communication lines, we use the recently suggested one-way scheme of key distribution with fast polarization encoding~\cite{Duplinskiy2017}. 
The setup is based on LiNbO3 phase modulators, single laser source for states generation, and two single-photon detectors (see Fig.~\ref{fig:setup}).  
An important improvement in compare with recent experiments on realizing three-node QKD network in Moscow~\cite{Pozhar2017} 
is the inclusion of an intensity modulator to the optical scheme as well as updating control units and post-processing software for the implementation of the decoy-state QKD protocol.
Quantum-generated keys then used for continuous key renewal in the hardware devices for establishing quantum-secured VPN Tunnel by Amicon~\cite{Amicon}.
The used fiber-optic communication lines are deployed between the Sberbank office on Bol'shaya Andron'yevskaya street (Alice) and the Sberbank office on Vavilova street (Bob): 
the one is used for QKD and another one for information transmitting.

\section{Experiment}

The optical scheme (Fig.~\ref{fig:setup}) realizing decoy-state BB84 QKD works as follows~\cite{Duplinskiy2017}.  
The laser source (L1) emits polarized optical pulses at 1550 nm. 
Then half-wave plate transforms the polarization state so that the amplitudes along the crystal axes of Alice’s phase modulator (PM 1) are equal to each other. 
This allows Alice to encode bits of the secret key in polarization states with the help of the modulator. 
To weaken the pulse, a variable optical attenuator (VOA) is used. 
After the quantum channel (QC), the piezo-driven polarization controller (PC) compensates SOP (state of polarization ) drifts 
and rotates it so that the polarization components along the lithium niobate crystal axes switch places, compensating the birefringence of LiNbO3. 
Bob's modulator PM 2 is used for basis selection. 
Finally, a half-wave plate ($\lambda/2$) converts SOP for polarization beam splitter (PBS) to distinguish states with the help of single-photon detectors (SPD1, SPD2). 
The decoy-state QKD protocol is realized by using intensity modulators. 
Polarization recalibration is applied once quantum bit error rate (QBER) in decoy pulses rises above the 8\% value. 
Then the gradient descent algorithm is applied for polarization controller to minimize the QBER. 
As soon as QBER over all types of pulses is under 5.5\%, the calibration is over and key generation is restarted.

The parameters of the QKD setup implementation are as follows: 
number of pulses in train $9.82\times10^4$, 
repetition rate of pulses in train 312,5 MHz, 
detectors efficiencies are 10\% and 6.4\% (for SPD1 and SPD2, respectively; see Fig.~1), 
detectors dead time 5~$\mu s$, 
dark count probability $3\times10^{-7}$, 
fiber channel losses 14.05~dB in the channel of 25 km length  (which corresponds to $\approx70$km of standard fiber-optic communication line with 0.2dB/km losses),
and additional losses on Bob's side 6~dB.
The communication line between two server rooms consists of 8 segments (6 segments outside the buildings and 2 inside the buildings). 
Few connections give us about $\approx$4\% of reflection. 
Toward to prevent the detector blinding, we separate clock synchronisation signal and quantum signal not only in wavelength but also on time.
The resulting raw key generation rate in our experiments is $\approx$2 kbit/s. 
After realization of the QKD session, we realize the standard sifting procedure, 
which is needed for dropping the positions with inconsistent bases from the raw quantum keys, 
by using authenticated communication channel (see below). 
The resulting keys are called sifted keys. 
The decoy states statistics~\cite{Trushechkin2017} is announced on this stage as well.

\begin{figure}[t]
\begin{centering}
\includegraphics[width=0.965\columnwidth]{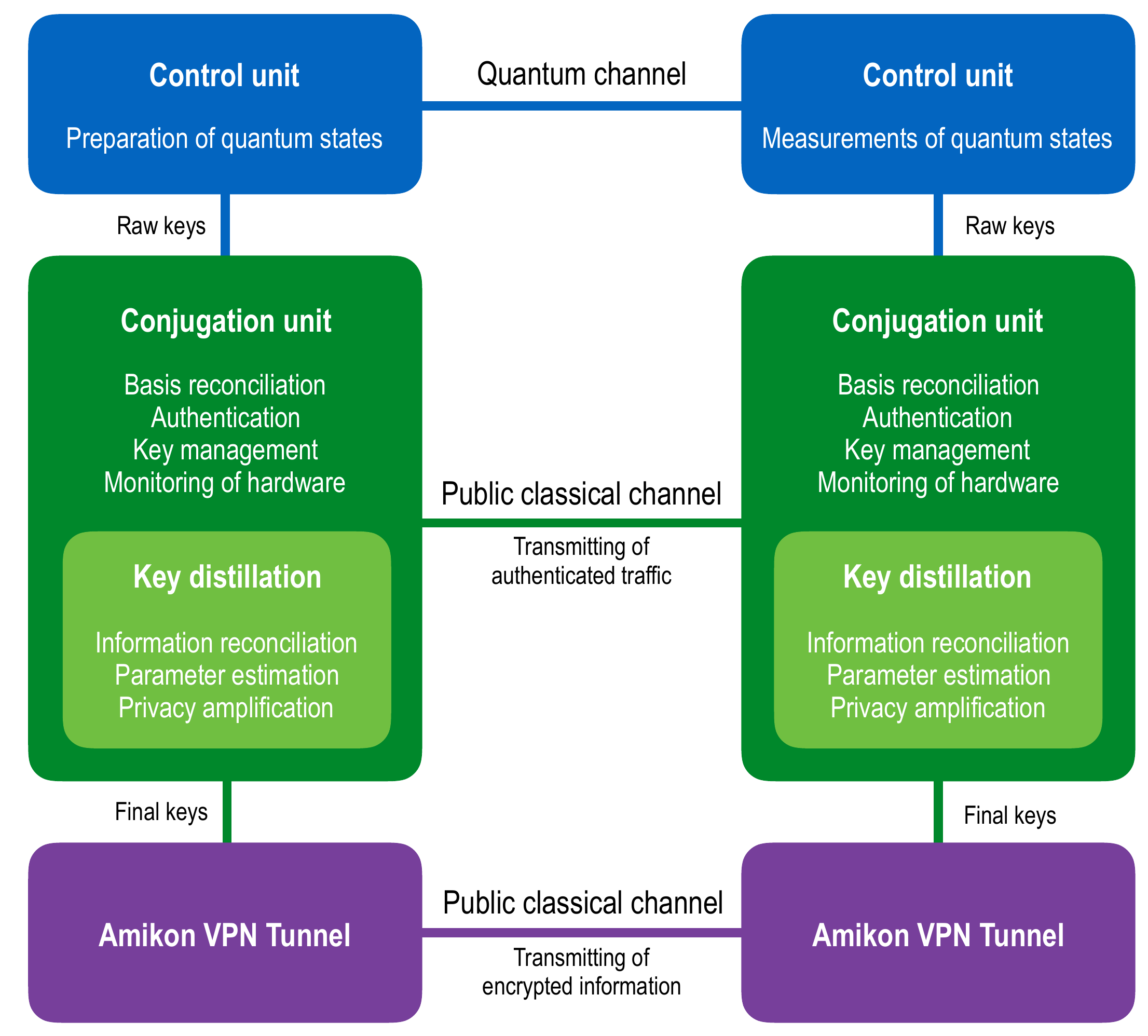}
\end{centering}
\vskip -4mm
\caption
{QKD technology stack, where control units realize the decoy-state QKD protocol. 
As a result of their work, raw quantum keys go to the basis reconciliation and to the post-processing procedures realized on conjugation units. 
After these stages, final secret keys can be requested by Amicon devices for establishing VPN tunnels.}
\label{fig:stack}
\end{figure}

\section{Post-Processing Procedure and Application Level}

The sifted keys are the input for a post-processing procedure~\cite{Kiktenko2016}.
The post-processing procedure includes a number of stages: information reconciliation, parameter estimation, privacy amplification, and, finally, authentication check. 
First, sifted keys from the hardware devices go through the information reconciliation stage.
We use the recently suggested symmetric blind information reconciliation method~\cite{Kiktenko2017}.
It uses low-density parity-check (LDPC) codes with frame length $l_{\rm frame}=4000$.
For a coarse tuning of the code rate we employ a pool of LDPC codes consisting of nine codes with the following rates: $\mathcal{R}=\{0.9,0.85,\ldots,0.55,0.5\}$.
For a fine tuning of the code rate, we employ the shortening and puncturing techniques~\cite{Elkouss2010,Elkouss2011}.\,We note that the total number of shortening and puncturing bits
was kept at constant level as follows:
\begin{equation}
	l_{\rm sp}=0.05l_{\rm frame}=200 \mbox{ bits}.
\end{equation}
The sub-block length of sifted key processed in a single launch of the symmetric blind reconciliation is as follows:
\begin{equation}
	l_{\rm sift} = l_{\rm frame}-l_{\rm sp}=3800 \mbox{ bits}.
\end{equation}
Here, $N_{\rm sift}=100$ sub-blocks of sifted keys were processed in parallel launches of the symmetric blind reconciliation method.
The resulting length of the sifted key processed in one round of the post-processing procedure was $L_{\rm sift}=380000$ bits.

After performing the information reconciliation stage, there is still a certain probability that uncorrected errors remain. 
In order to detect possible remaining errors, we use the subsequent verification protocol with the use of $\epsilon$-universal hash functions~\cite{Kiktenko2017_2}.
The probability of the presence of errors after successful verification of the block of $N_{\rm sift}$ bits is bounded by the value of $\epsilon_{\rm ver}=1.4\times10^{-11}$ 
with the use of a hash-tag of 50 bit length.
Due to the low level of frame error rate of the employed LDPC codes, 
we obtain the length of verified keys $L_{\rm ver}$ to be almost always equal to the length of the processed sifted keys ${L_{\rm ver}}\approx L_{\rm sift}$.

The next stage in the post processing is the parameter estimation stage. 
On this stage, the parties obtain the actual level of the QBER $q$ for their key blocks via direct comparison of the keys before and after the information reconciliation.
If the value of QBER appeared to be higher than the critical value needed for efficient privacy amplification (11\% for the decoy-state BB84 protocol), 
the parties receive a warning message about possible eavesdropping. 
Otherwise, the verified key blocks go to the privacy amplification stage, and estimated QBER is used in next rounds of the information reconciliation stage. 
In our experiments, QBER was on the level of 4.8\%-6\%, so we were able successfully implement the privacy amplification procedure.

The aim of the privacy amplification stage is to reduce potential information of an adversary about the verified blocks to a negligible quantity~\cite{Gisin2002}. 
Such a reduction can be achieved by a contraction of the input verified key into a shorter key. 
The length of the secret key is computed as follows:
\begin{equation} \label{eq:pa}
	L_{\rm sec}=L_{\rm ver}\hat{Y}_1(1-h(\hat{q}_1))-{\rm leak}_{\rm ec}-5\log_2(1/\epsilon_{\rm pa}),
\end{equation}
where $L_{\rm ver}$ is length of the verified key, 
$\hat{Y}_1$ is an estimation of the portion of the sifted key bits generated from single photons pulses,
$$
	h(x)=-x\log_2 x-(1-x)\log_2(1-x)
$$
is binary entropy function, 
$\hat{q}_1$ is an estimation of the QBER for single photon pulses, 
${\rm leak}_{\rm ec}$ is total number of bits disclosed in information reconciliation and verification stages, 
and $\epsilon_{\rm pa}$ is the failure probability of privacy amplification stages ($\epsilon_{\rm pa}=10^{-12}$ in our setup).

The estimates of $\hat{Y}_1$ and $\hat{q}_1$ were obtained using the decoy-states method.
We employed three types of pulses with different intensities $\mu\approx0.175$ (signal), $\nu\approx0.067$ (decoy), and $\lambda\approx0.008$ (vacuum).
The corresponding probabilities of generating each type of pulses were as follows: $p_\mu=0.5$, $p_\nu=p_\lambda=0.25$.
We note that the sifted key was generated using signal pulses only~\cite{Trushechkin2017}.
The length of the secret key can be then calculated as a function of the following form:
\begin{equation}
	L_{\rm sec}=L_{\rm sec}(L_{\rm ver}, \mu, \nu, \lambda, N_\mu, n_\mu, N_\nu, n_\nu, N_\lambda, n_\lambda, q),
\end{equation}
where $N_x$ and $n_x$ are the numbers of sent and detected states of intensity $x\in\{\mu, \nu, \lambda\}$.
The detailed description of the function $L_{\rm sec}$ can be found in Ref.~\cite{Trushechkin2017}.

\begin{figure}[t]
\begin{centering}
\includegraphics[width=1\columnwidth]{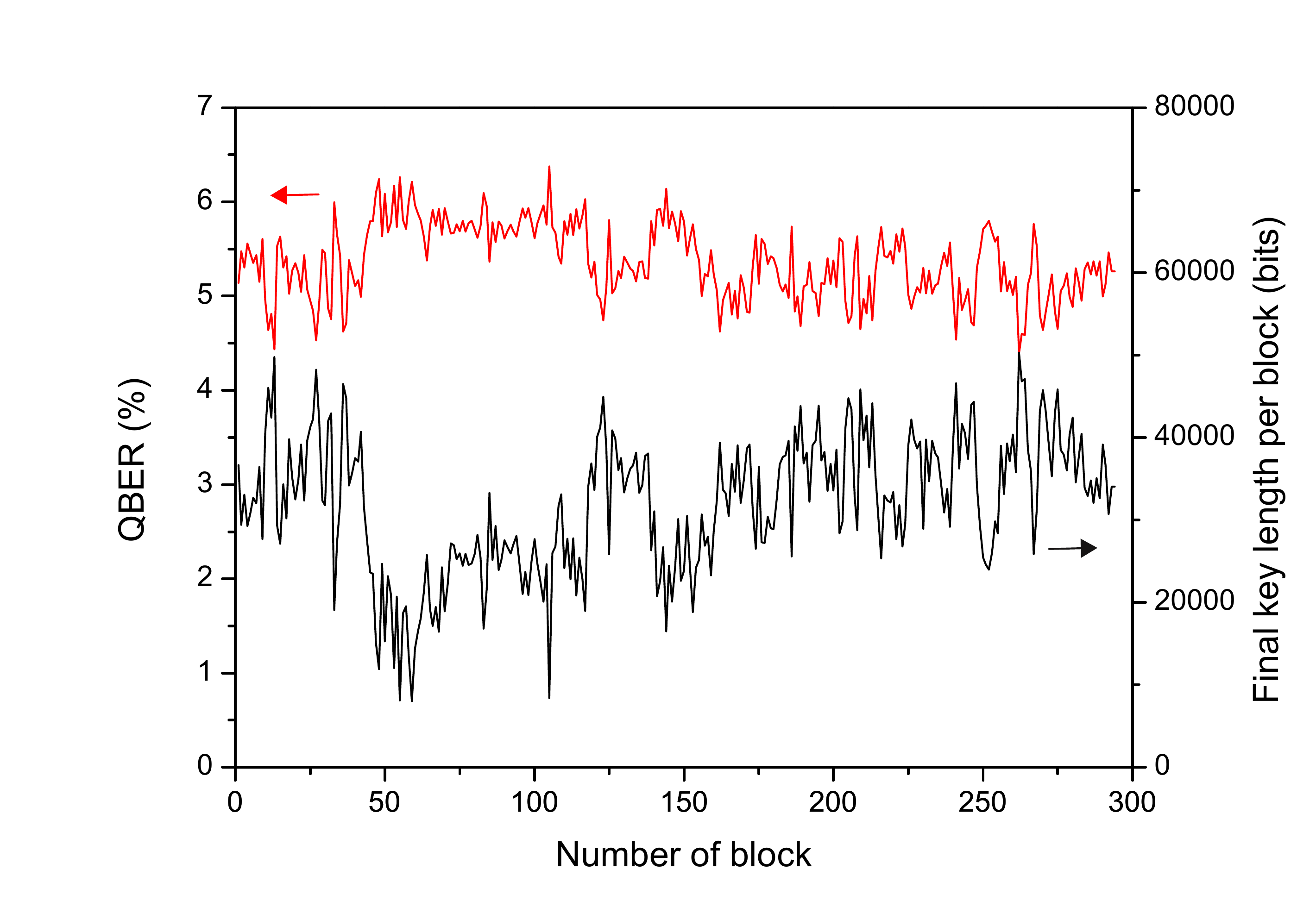}
\end{centering}
\vskip -8mm
\caption
{QBER (upper curve) and the length of final keys (lower curve) are shown as functions of the generated block indices.
Each block of the final key is obtained from a block of the sifted keys of $L_{\rm sift} = 380000$ bits length.}
\label{fig:qber}
\end{figure}

After the calculation of the length of the secret key, the privacy amplification can be realized. 
On this step, the block of the secret key is computed as a result of application of the 2-universal hash function to the verified key~\cite{Kiktenko2016}.  
In our setup the Toeplitz hashing is used.

At the final state, the parties need to check the authenticity of their communications over the classical channel by an exchange of hash values of the whole incoming traffic.
For this purpose, we employ the information-theoretically secure Toeplitz hashing together with one time-pad encryption~\cite{Kiktenko2016}.
The length of the hash value is $l_{\rm auth}=40$ bit, which bounds the probability of successful man in the middle attack at the level of
\begin{equation}
	\epsilon_{\rm auth}=2\times 2^{-l_{\rm auth}}<2\times 10^{-12}.
\end{equation}

If the authenticity is verified, the parties reserve $2l_{\rm auth}$ bits of their secret quantum keys for the next post-processing round and obtain
\begin{equation}
	L_{\rm fin} = L_{\rm sec}-2l_{\rm auth}
\end{equation}
bits of the final key that can be used in cryptographic purposes.
We then obtain the length of the secret key $L_{\rm fin}$ to be $0.03\ldots0.13$ of $L_{\rm ver}$ depending on QBER (see Fig.~\ref{fig:qber}).
The final security level of the obtained key is given by
\begin{equation}
	\epsilon_{\rm QKD}=\epsilon_{\rm ver}+\epsilon_{\rm pa}+\epsilon_{\rm auth}<2\times 10^{-11}.
\end{equation}
As a result, after the post-processing procedure, from $2$ kbit/s of sifted keys, we obtain about $0.1$ kbit/s of secret keys. 
This value can be improved significantly by fine tuning of the parameters of the decoy-state QKD protocol, 
stabilization of the hardware, and improving characteristics of the fiber-optic communication line. 

After post processing, quantum-generated keys are used for continuous key renegotiation in the hardware devices for establishing quantum-secured VPN Tunnel.
The VPN Tunnel performs L3-level encryption using the Russian symmetric block cipher algorithm (GOST 28147-89) with a 256 bit key size.
In our experimental tests, hardware device establishing the VPN Tunnel was connected to the QKD setup via the Ethernet channel. 
Using the special API-protocol the VPN Tunnel device requests a new quantum key every 400 seconds, which adds to the master keys of the device. 
In the case of successful obtaining symmetric quantum-generated keys on the both sides, then encryption of transmitted data is performed using both session and quantum keys, i.e. a hybrid scheme.
Data transfer rate in the hybrid encryption scheme is about 1 Gbit/s. 
Up to our knowledge, this is a first in Russia experimental demonstration of quantum-secured data transmission in urban fibre-optic communication lines, 
while previously announced results were about implementations of QKD protocols only~\cite{Balygin2017, Glem2017}.

\section{Conclusions}

QKD technology provides the ultimate in quantum-safe security, 
guaranteeing provably secure key exchange for encryption and other security devices on point-to-point backbone, networks, and distributed ledgers, 
such as blockchains~\cite{Kiktenko2017_3}.
We emphasize that the realized hybrid approach, where quantum-generated keys are used for continuous key renewal in already existing information security solutions,  
offers the method for long-term data protection in the post-quantum era.
Furthermore, we expect that using a high-quality fiber-optic communication line (e.g. with 0.2 dB/km loss coefficient) and improving all stabilization issues in hardware and software results in 
an increase of the key generation rate up to 100 kbit/s, which is enough for transmitting audio information in the one-time pad regime. 

{\bf Acknowledgments}.
We express our gratitude to Mr. S. V. Lebed’, the head of Cybersecurity Division of the Sberbank and Mr. S. K. Kuznetsov, the Deputy Chairman of the Sberbank Board, 
as well as the colleagues from Amicon for their help in realizing this experimental work. The work was supported by the Russian Science Foundation under Grant No. 17-71-20146.


\begin{thebibliography}{}

\bibitem{Schneier}
B. Schneier,
{\it Applied cryptography} 
(John Wiley \& Sons, Inc., New York, 1996).

\bibitem{Shor1997}
P.W. Shor,
{\href{https://doi.org/10.1137/S0036144598347011}{SIAM J. Comput. {\bf 26}, 1484 (1997)}}.

\bibitem{Grover1996}
L.K. Grover,
in {\it Proceedings of 28th Annual ACM Symposium on the Theory of Computing (New York, USA, 1996)}, p. 212.

\bibitem{Gisin2002}
N. Gisin, G. Ribordy, W. Tittel, and H. Zbinden,
{\href{https://dx.doi.org/10.1103/RevModPhys.74.145}{Rev. Mod. Phys. {\bf 74}, 145 (2002)}}.

\bibitem{Lo2015}
H.-K. Lo, M. Curty, and K. Tamaki,
{\href{https://dx.doi.org/10.1038/nphoton.2014.149}{Nat. Photonics {\bf 8}, 595 (2014)}}.

\bibitem{Lo2016}
E. Diamanti, H.-K. Lo, and Z. Yuan, 
{\href{https://dx.doi.org/10.1038/npjqi.2016.25}{npj Quant. Inf. {\bf 2}, 16025 (2016)}}.

\bibitem{Hwang2003}
W.-Y. Hwang, 
{\href{https://doi.org/10.1103/PhysRevLett.91.057901}{Phys. Rev. Lett. {\bf 91}, 057901 (2003)}}.

\bibitem{Lo2005}
H.-K. Lo, X. Ma, and K. Chen, 
{\href{https://doi.org/10.1103/PhysRevLett.94.230504}{Phys. Rev. Lett. {\bf 94}, 230504 (2005)}}.

\bibitem{Wang2005}
X.-B. Wang, 
{\href{http://dx.doi.org/10.1103/PhysRevLett.85.1330}{Phys. Rev. Lett. {\bf 94}, 230503 (2005)}}.

\bibitem{Ma2005}
X. Ma, B. Qi, Y. Zhao, and H.-K. Lo,
{\href{https://doi.org/10.1103/PhysRevA.72.012326}{Phys. Rev. A {\bf 72}, 012326 (2005)}}.

\bibitem{Curty2014}
M. Curty, F. Xu, W. Cui, C.C.W. Lim, K. Tamaki, and H.-K. Lo, 
{\href{https://doi.org/10.1038/ncomms4732}{Nat. Commun. {\bf 5}, 3732 (2014)}}.

\bibitem{Lim2014}
C.C.W. Lim, M. Curty, N. Walenta, F. Xu, and H. Zbinden, 
{\href{https://doi.org/10.1103/PhysRevA.89.022307}{Phys. Rev. A {\bf 89}, 022307 (2014)}}.

\bibitem{Ma2017}
Z. Zhang, Q. Zhao, M. Razavi, and X. Ma,
{\href{https://doi.org/10.1103/PhysRevA.95.012333}{Phys. Rev. A {\bf 95}, 012333 (2017)}}.

\bibitem{Trushechkin2017}
A.S. Trushechkin, E.O. Kiktenko, and A.K. Fedorov,
{\href{https://doi.org/10.1103/PhysRevA.96.022316}{Phys. Rev. A {\bf96}, 022316 (2017)}}.

\bibitem{Duplinskiy2017}
A. Duplinskiy, V. Ustimchik, A. Kanapin, V. Kurochkin, and Y. Kurochkin, 
{\href{https://dx.doi.org/10.1364/OE.25.028886}{Opt. Express {\bf 25}, 28886 (2017)}}.

\bibitem{Pozhar2017}
E.O. Kiktenko, N.O. Pozhar, A.V. Duplinskiy, A.A. Kanapin, A.S. Sokolov, S.S. Vorobey, A.V. Miller, V.E. Ustimchik, M.N. Anufriev, 
A.S. Trushechkin, R.R. Yunusov, V.L. Kurochkin, Y.V. Kurochkin, and A.K. Fedorov,
{\href{http://dx.doi.org/10.1070/QEL16469}{Quantum Electron. {\bf 47}, 798 (2017)}}.

\bibitem{Amicon}
Web-site: {\href{https://www.amicon.ru/page.php?link=fpsu-ip}{Amicon FSPU-IP}}. 

\bibitem{Kiktenko2016}
E.O. Kiktenko, A.S. Trushechkin, Y.V. Kurochkin, and A.K. Fedorov,
{\href{http://dx.doi.org/10.1088/1742-6596/741/1/012081}{J. Phys. Conf. Ser. {\bf 741}, 012081 (2016)}}.

\bibitem{Kiktenko2017}
E.O. Kiktenko, A.S. Trushechkin, C.C.W. Lim, Y.V. Kurochkin, and A.K. Fedorov,
{\href{https://doi.org/10.1103/PhysRevApplied.8.044017}{Phys. Rev. Applied {\bf 8}, 044017 (2017)}}.

\bibitem{Elkouss2010}
D. Elkouss, J. Mart{\'{\i}}nez{-}Mateo, and V. Martin,
Secure rate-adaptive reconciliation,
{\href{https://dx.doi.org/10.1109/ISITA.2010.5650099}
{in {\it Proceedings of the IEEE International Symposium on Information Theory and its Applications (ISITA)}, Taichung, Taiwan (IEEE, 2010), p. 179}}.

\bibitem{Elkouss2011}
D. Elkouss, J. Mart{\'{\i}}nez{-}Mateo, and V. Martin,
Information reconciliation for quantum key distribution,
{\href{https://dblp2.uni-trier.de/db/journals/qic/qic11.html}{Quant. Inf. Comp. {\bf 11}, 226 (2011)}}.

\bibitem{Kiktenko2017_2}
A.S. Trushechkin, E.O. Kiktenko, and A.K. Fedorov,
{\href{https://arxiv.org/abs/1705.06664}{arXiv:1705.06664 (2017)}}.

\bibitem{Balygin2017}
K.A. Balygin, V. I. Zaitsev, A.N. Klimov, A.I. Klimov, S.P. Kulik, and S.N. Molotkov,
{\href{https://doi.org/10.1134/S0021364017090077}{JETP Lett. {\bf 105}, 606 (2017)}}.

\bibitem{Glem2017}
A.V. Gleim, V.V. Chistyakov, O.I. Bannik, V.I. Egorov, N.V. Buldakov, A.B. Vasilev, A.A. Gaidash, A.V. Kozubov, S.V. Smirnov, S.M. Kynev, S.E. Khoruzhnikov, S.A. Kozlov, and V.N. Vasil'ev,
{\href{https://doi.org/10.1364/JOT.84.000362}{J. Opt. Tech. {\bf 84}, 362 (2017)}}.

\bibitem{Kiktenko2017_3}
E.O. Kiktenko, N.O. Pozhar, M.N. Anufriev, A.S. Trushechkin, R.R. Yunusov, Y.V. Kurochkin, A.I. Lvovsky, and A.K. Fedorov,
{\href{https://arxiv.org/abs/1705.09258}{arXiv:1705.09258 (2017)}}.

\end{thebibliography}
\end{document}